# Tunneling of trapped-atom Bose condensates

SUBODH R SHENOY

Abdus Salam International Centre for Theoretical Physics, P.O. Box 586, Trieste 34100, Italy

**Abstract.** We obtain the dynamics in number and phase difference, for Bose condensates that tunnel between two wells of a double-well atomic trap, using the (nonlinear) Gross–Pitaevskii equation. The dynamical equations are of the canonical form for the two conjugate variables, and the Hamiltonian corresponds to that of a *momentum-shortened* pendulum, supporting a richer set of tunneling oscillation modes than for a superconductor Josephson junction, that has a fixed-length pendulum as a mechanical model. Novel modes include 'inverted pendulum' oscillations with an average angle of $\pi$; and oscillations about a self-maintained population imbalance that we term 'macroscopic quantum self-trapping'. Other systems with this phase-number nonlinear dynamics include two-component (interconverting) condensates in a single harmonic trap, and $He^3B$ superfluids in two containers connected by micropores.

## 1. Introduction

In 1995 alkali atoms in magnetic traps, cooled to a tenth of a microkelvin by laser and evaporative methods, were shown to undergo Bose–Einstein condensation to produce a common condensate wave function [1]. This collective quantum state has a collective quantum phase, as indicated by atom interference fringes formed when two falling Bose–Einstein condensates (BEC) begin to overlap [2]. In a theoretical and experimental search for BEC analogues of superfluid coherence, quantized vortex lines [3] or quantized rotational superflow in toroidal-geometry traps [4–6] have been investigated. The Josephson tunneling of Cooper pairs through superconductor junctions is another analogue to explore, in trapped atom systems [7–17]. This coherence phenomenon is especially interesting because it would involve tunneling through energy barriers of whole atoms, and not just paired electrons. Reviews of the rich physics exhibited by trapped atoms are available [18], and we will not discuss all the Josephson tunneling literature but will focus on work done at Trieste.

We will summarize the essential ideas of a series of theoretical papers [10–14] on tunneling of one-component Bose condensates between double-well traps; between two-component condensates in a single harmonic trap; and between $He^3B$ superfluids connected by narrow pores. We predict that boson Josephson junctions (BJJ) of neutral atoms exhibit not only the standard tunneling modes of superconductor Josephson junctions (SJJ), but also novel modes that have no superconductor analogue. While the SJJ has a well-known mechanical model of a rigid pendulum [19], the BJJ maps onto a *momentum-shortened* pendulum that gives an intuitive understanding of this richer set of modes.

This meeting is (mostly) on the physics of metallic mesoscopic systems, with electrons as carriers, that have large impurity scattering and thermal dephasing lengths, comparable to sub-micron-scale system sizes. This regime allows an investigation of interacting-electron coherent transport at low temperatures, and quantum effects of multiply connected wire and junction geometries on propagation and tunneling [20]. A normal-state elliptical-trap atom system of around a hundred thousand bosons, or even of degenerate fermions [21], is of course, not quite a metallic wire. However, such a system is on a scale of tens of microns, and at sub-microkelvin temperatures. It has no quenched disorder, but has the possibility of added 'impurity'-atom annealed disorder. The magnetic or optical trap potentials of the system directly control coherent wave functions, with trap geometries that can be quasi-1D cigars [22], or nested circular rings [4], or multiwell arrays [16]. Furthermore, rotating flows can mimic magnetic fields; tilted traps with micron height differences can induce gravitational potential energies of tens of nanokelvin that are like voltages; trap potentials and barriers can have added random components fluctuating in space and time; quantum fluctuations can occur for small numbers of trapped atoms; and the interatomic interaction can be tunable in magnitude and sign [23]. These interesting results suggest that although the detailed physics will be different, ideas from electronic mesoscopics might be usefully examined in the context of multiply connected atom traps and junctions, for developing an *atomic* mesoscopics of both bosons and fermions, and in both normal and superfluid states.

However, here we will focus on the simpler problem of the two-state BEC tunneling. In §2 we outline the derivation of the double-well BJJ tunneling equations, and compare them to the SJJ dynamics. In §3 we present the momentum-shortened pendulum model. Section 4 outlines other systems, and §5 is a conclusion.

## 2. Dynamics of a boson Josephson junction

Bose–Einstein condensates in a trap $V(\vec{r})$ [1] are well-described by the Gross–Pitaevskii equation (GPE) [18] for the condensate wave function $\Psi(\vec{r},t)$

$$i\hbar\dot{\Psi} = -[\hbar^2\nabla^2/2M]\Psi + V\Psi + g|\Psi|^2\Psi, \tag{1}$$

where $g = 4\pi\hbar^2 a/M$, and $a$ is the (positive) s-wave atom–atom scattering length while $M$ is the atomic mass. We assume $V(\vec{r})$ is an effective double-well trap, with wells $1,2$ on either side, for $x<0$, $x>0$, of a narrow barrier along the y-axis. (This can be done e.g. with an off-resonant laser beam bisecting a harmonic magnetic trap.) The trap contains a fixed total number of atoms $N_1(t) + N_2(t) = N_\text{T}$. We use the stationary solutions of the nonlinear GPE, $\Phi_{1,2}(\vec{r})$, that depend on the particle number, with $N_\text{T}/2$ particles in each well [10,11]. (One can think of $\Phi_{1,2}$ as combinations of even and odd states in the full double well.) The solutions are taken to be unit-normalized, and have negligible overlap:

$$\int d^3r |\Phi_{1,2}(\vec{r})|^2 = 1, \tag{2}$$

$$\int d^3r \Phi_1(\vec{r})\Phi_2^*(\vec{r}) \ll 1. \tag{3}$$

The time-dependent condensate wave function is then assumed to be a linear combination of these rigid-profile wave functions peaked at potential minima, but now scaled up or down in the number of particles by time-dependent amplitudes $\psi_{1,2}(t)$ [10,11]. Thus

$$\Psi(\vec{r},t) = \psi_1(t)\Phi_1(\vec{r}) + \psi_2(t)\Phi_2(\vec{r}), \tag{4}$$

where the amplitudes are

$$\psi_{1,2}(t) = \sqrt{N_{1,2}(t)}e^{i\phi_{1,2}(t)}. \tag{5}$$

It is convenient to define the relative phase $\phi(t)$ and the fractional number imbalance $\eta(t)$ by

$$\phi(t) = \phi_2(t) - \phi_1(t), \tag{6}$$

$$\eta(t) = [N_1(t) - N_2(t)]/N_T. \tag{7}$$

Substituting the ansatz of (4) into (1) and integrating over space gives the amplitude equations [12]

$$i\hbar\dot{\psi}_{1,2} = E_{1,2}\psi_{1,2} + U_{1,2}|\psi_{1,2}|^2\psi_{1,2} - K\psi_{2,1}, \tag{8}$$

where $E_{1,2}, U_{1,2}$ and $K$ are respectively the chemical potentials in the wells, the interaction energies $\propto a$, and the interwell coupling. Note that these parameters are matrix-element integrals over $\Phi_{1,2}$, and so depend on $N_T$.

The dynamic equations for the conjugate variables (relative phase and relative number) then follow immediately, giving a two-state description of the BJJ [10,11]:

$$\tau\dot{\eta} = -\sqrt{1-\eta^2}\sin\phi = -\partial H/\partial\phi, \tag{9}$$

$$\tau\dot{\phi} = \Lambda\eta + \Delta E + \eta\cos\phi/\sqrt{1-\eta^2} = \partial H/\partial\eta. \tag{10}$$

The BJJ tunneling current is $I = \tau^{-1}N_T\sqrt{1-\eta^2}\sin\phi$. The intrinsic inverse time scale $\tau^{-1} = 2K/\hbar$ is typically of the order of kilohertz for the BJJ, in contrast to the SJJ where the basic frequency scale is gigahertz. The (dimensionless) Hamiltonian is

$$H = \Lambda\eta^2/2 + \Delta E\eta - \sqrt{1-\eta^2}\cos\phi, \tag{11}$$

where $\Delta E = (E_1 - E_2)/2K + (U_1 - U_2)/4K$, and the interaction parameter is $\Lambda = UN_T/2K$, with $U = (U_1 + U_2)/2$.

The time-dependent nonlinear equations can be solved exactly in terms of the elliptic functions [11]. The small-amplitude oscillations around $\phi = \eta = 0$ have frequency

$$\omega_0 = \sqrt{1+\Lambda}/\tau. \tag{12}$$

The BJJ equations differ from the usual dynamics for the superconductor Josephson junction, that in this notation can be written as [24]

*Subodh R Shenoy*

$$\tau\dot{\eta} = -\sin\phi = -\partial H/\partial\phi, \tag{13}$$

$$\tau\dot{\phi} = \Delta\mu = \partial H/\partial\eta, \tag{14}$$

where the chemical potential difference is $\Delta\mu = \Delta E + \Lambda\eta$ and $\Lambda \propto E_c$, the junction charging energy per Cooper pair. The first (eq. (13)) is the familiar expression for the Josephson current $I \propto \sin\phi$, and the second (eq. (14)) is the Josephson phase relation. The SJJ Hamiltonian is

$$H = \Lambda\eta^2/2 + \Delta E\eta - \cos\phi. \tag{15}$$

Equations (13) and (14) support small-amplitude Josephson 'plasma' oscillations of frequency

$$\omega_J = \sqrt{\Lambda}/\tau \tag{16}$$

proportional to $\sqrt{E_J E_c}$, the geometric mean of the Josephson coupling and the charging energy.

Feynman has given a simple two-state phenomenological model [24] for Cooper-pair amplitudes $\psi_{1,2} = \sqrt{N_{1,2}}e^{i\phi_{1,2}}$ to describe Josephson tunneling in superconductors through linear, Schrödinger-like equations

$$i\hbar\dot{\psi}_{1,2}(t) = E_{1,2}\psi_{1,2} - K\psi_{2,1}, \tag{17}$$

which are like two-level atom amplitudes, with $K$ the transition matrix element. They are also the noninteracting $\Lambda = 0$ limit of the GPE-derived equations (8). The (phenomenological) Feynman tunneling current is $I \propto K\sqrt{N_1 N_2}\sin(\phi_2 - \phi_1)$, which is like the (microscopically derived) BJJ form. However, because the external circuit rapidly removes excess charges, the number of charged Cooper pairs $N_{1,2}$ in the superconductors on either side of the junction is essentially fixed. Even if we consider isolated superconductors separated by a junction, the requirement that developed capacitive voltages be less than twice the superconducting gap (to avoid boiling off pairs into quasiparticles) implies, with typical Josephson coupling and Coulombic charging energies, that the imbalance is completely negligible, $|\eta| < 10^{-9}$. Thus $N_{1,2}$ are essentially locked to their average values and the familiar Josephson relation $I \propto K\sin(\phi_2 - \phi_1)$ for the SJJ emerges. The SJJ charged-pair current is then seen to be a special limiting case of the more general, and generic, BJJ form, where $\eta$ is allowed to be appreciable for the neutral-atom case. Whereas the Josephson plasma frequency $\omega_J$ vanishes for uncharged Cooper pairs, the 'zero-state' BJJ frequency $\omega_0$ interpolates between an ideal Bose gas *nonzero* value of $1/\tau$ for $\Lambda = 0$, and the SJJ form of $\sqrt{\Lambda}/\tau$ for $\Lambda \gg 1$. Quantum effects can be ignored in the SJJ for $E_c/E_J \ll 1$ [24], and since $E_c, E_J$ correspond to $U, KN_T$ in the BJJ case, a similar argument yields the wide regime of validity of the semiclassical BJJ dynamics, as $\Lambda(N_T) < N_T^2$.

We note that if we linearize (9) and (10) in $\eta$ only, then the resulting ($\Delta E = 0$) equation for the phase is $\tau^2\ddot{\phi} = -\partial V/\partial\phi$, where $V = -\Lambda\cos\phi - (1/4)\cos 2\phi$. This effective potential $V$ plotted versus $\phi/\pi$ for various $\Lambda$ [11] has, in addition to the familiar even-integer minima, a new series of metastable minima at odd-integers, implying possible new oscillation modes.

## 3. Tunneling modes and pendulum models

The superconducting Josephson junction is well-known [19] to have a mechanical model: eq. (15) is the Hamiltonian of a pendulum of fixed (unit) length, tilt angle $\phi$, (dimensionless) angular momentum $p_\phi = \eta$, inverse mass $\Lambda$, and applied torque $\Delta E$. The harmonic frequency (16) is that of a pendulum oscillating about its only stable fixed point $p_\phi^* = \phi^* = 0$, a downward oriented pendulum at rest. The pendulum can also rotate, with a nonzero average angular momentum and a freely running phase.

The BJJ has, in addition to such modes, a more complex oscillation behavior [10,11]. The Hamiltonian of (11) corresponds to a momentum-shortened pendulum of length $\sqrt{1-p_\phi^2}$: 'faster equals shorter', so for small oscillations about equilibrium, the length is least at the bottom of the swing [12]. Note that the length depends on the canonical *momentum*, and the angle is still the single canonical *coordinate*. (This is unlike the pendulum with an elastic string that has both the angle and the string extension as two canonical coordinates. However, see later discussion of BEC profile collective modes.)

The BJJ dynamical equations have several fixed points $\eta^*, \phi^*$ and linearizing in the deviations yields the harmonic frequencies. Five possible modes can occur, which are illustrated in figures 1a–d, through numerical solutions of the BJJ dynamics, with initial conditions $\eta(0) = 0.25, \phi(0) = 0$ for figures 1a,b; and with $\eta(0) = 0.75, \phi(0) = \pi$ for figures 1c,d. The $\Lambda$ parameters used to illustrate the various modes for figures 1a–d are chosen to be 5.0, 75.0, 0.35 and 2.0, respectively. Here, the pendulum cartesian coordinates are $x = \sqrt{1-\eta^2}\sin\phi$, $y = -\sqrt{1-\eta^2}\cos\phi$. The figures 1a–d pendulum trajectories correspond to:

(a) 'zero-state oscillations' of angular frequency $\omega_0 = \sqrt{1+\Lambda}/\tau$ around $\phi^* = 0 = \eta^*$,

(b) 'running-states' with small oscillations about a mean phase that increases with time, $\langle\phi\rangle \propto t$. This state has nonzero (positive or negative) mean angular momentum $\langle p_\phi\rangle = \langle\eta\rangle$, with the pendulum executing closed-loop rotations enclosing its point of support,

(c) '$\pi$-state oscillations' of angular frequency $\omega_\pi = \sqrt{1-\Lambda}/\tau$ about $\phi^* = \pi, \eta^* = 0$. Here, since the pendulum moves fastest at the middle of its swing, it has the least length there, and so 'digs a hole for itself' in a dynamically stable 'inverted pendulum' state.

(d) '$\pi$-state rotations' of angular frequency $\omega_\pi^{\rm rot} = \sqrt{\Lambda^2-1}/\tau$ linearized around $\phi^* = \pi$ and a nonzero momentum $\eta^* = \pm\sqrt{1-\Lambda^{-2}}$. In this state the length-varying pendulum executes closed-loop rotations above its point of support (slowest at the top of the closed locus, and fastest at its bottom). There are two varieties of such $\pi$-state rotations, with average momentum magnitudes $|\langle\eta\rangle|$ which are respectively less or greater than the fixed point value, $|\eta^*|$, as seen below.

As the amplitude of the zero-state increases either through higher initial values of tilts/momenta or by increasing the inverse mass $\Lambda$, there is a transition as the pendulum reaches the vertical. At this critical energy, or critical $\Lambda$ [11], the oscillation frequency of the zero-state dips to zero. Symmetry breaking occurs, and figure 1a goes over to figure 1b, with an onset of pendulum rotations or running states with nonzero $\langle p_\phi\rangle$, that are either clockwise or anticlockwise. Similarly, for the pendulum doing a $\pi$-state oscillation, as $\Lambda$ increases the amplitude increases till a critical $\Lambda$ is reached [11], where the oscillation frequency of the $\pi$-state goes to zero. Symmetry breaking occurs, and figure 1c goes over to figure 1d, with an onset of $\pi$-state rotations with nonzero $\langle p_\phi\rangle$, that are either clockwise or anticlockwise. There can be a further transition to a second type of $\pi$-state rotation [11].

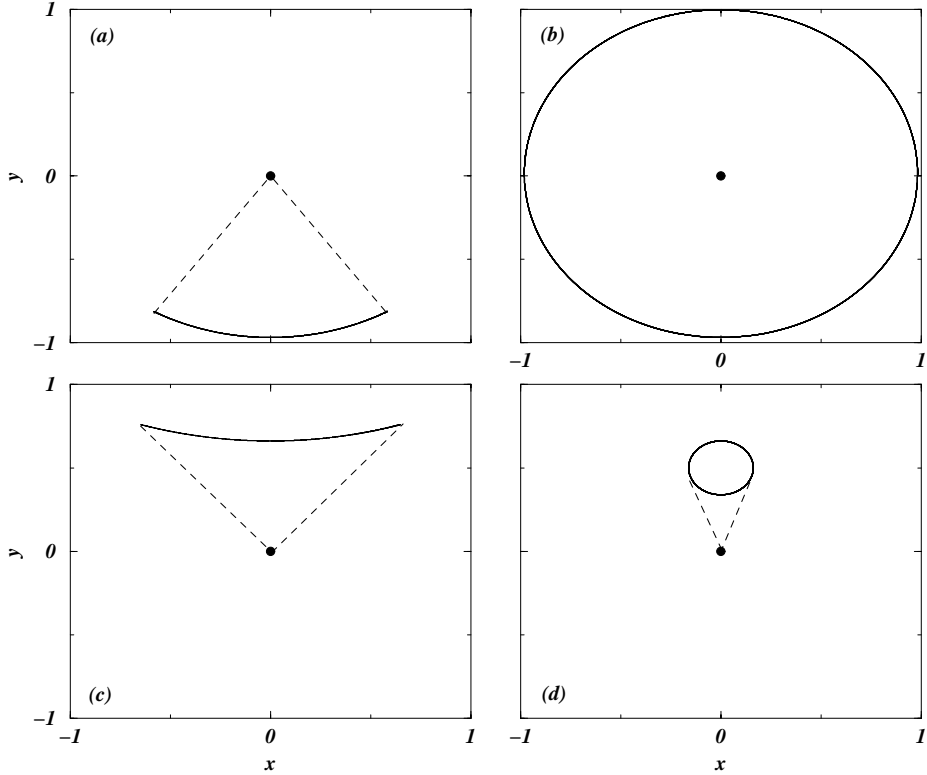

**Figure 1.** Trajectories of a 'momentum-shortened pendulum' from numerical solution of BJJ dynamic equations, with pendulum cartesian coordinates $(x,y) = (\sqrt{1-\eta^2}\sin\phi, -\sqrt{1-\eta^2}\cos\phi)$. (**a**) Zero-state oscillation; (**b**) running state rotation; (**c**) $\pi$-state oscillation; (**d**) $\pi$-state rotation of the first kind. A similar state of the second kind (see discussion of figure 2 in text) is qualitatively the same, and is not shown.

Note that for figures 1b and 1d, the nonzero average momentum $\langle p_\phi \rangle$ of the rotating states in the pendulum model corresponds in the physical BEC system to a nonzero average population imbalance $\langle \eta \rangle$, with more condensate atoms in one well than the other. An incremental number of atoms tunnel back and forth, causing small oscillations around this self-maintained number imbalance. We have termed this trapping phenomenon, arising from the nonlinear GPE equation for the macroscopic condensate wave function, as 'macroscopic quantum self-trapping' (MQST) [10,11]. Furthermore, the $\pi$-states here occur at the same order $K\cos\phi$ as the zero-states; with an *s*-wave condensate order parameter; and without any extraneous material to dope the barrier region. Our $\pi$-states therefore differ from those in SJJ, which can occur as second-order $K^2\cos^2\phi$ tunneling effects; or as *d*-wave order-parameter effects in high-temperature superconductor junctions [25]; or with ferromagnetic impurities in the oxide barrier [26].

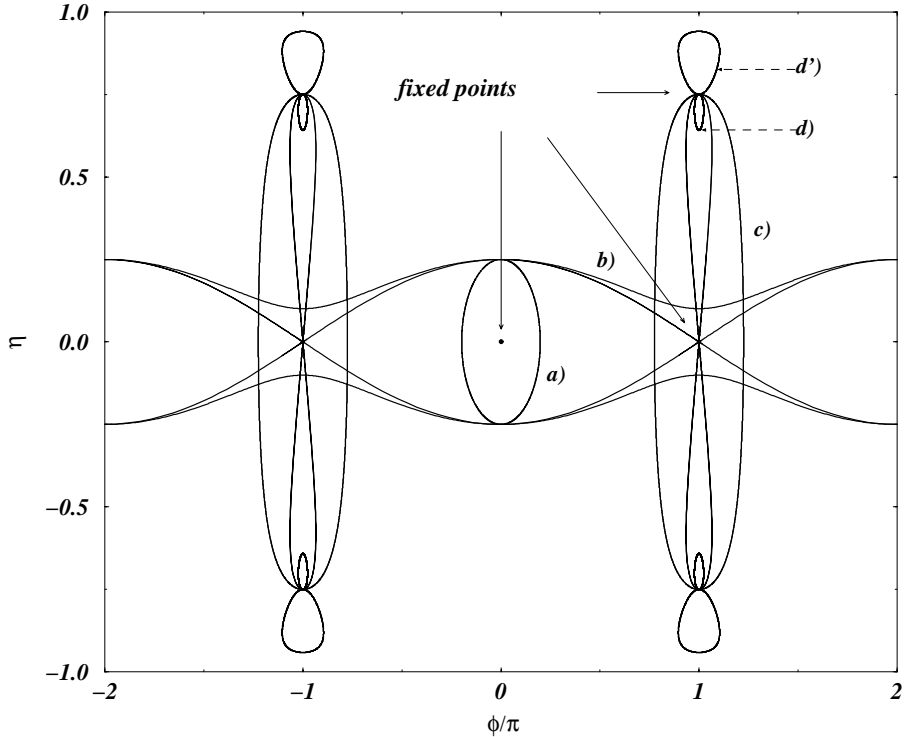

**Figure 2.** Phase space plot for the momentum-shortened pendulum. Vertical axis is the momentum $p_\phi = \eta$ versus scaled tilt angle $\phi/\pi$, showing the constant-energy trajectories labelled (a), (b), (c), (d) and (d$'$), which are discussed in the text. See also figures 1, 3.

Figure 2 shows a phase-space plot of the momentum-shortened pendulum, where the lines are the trajectories in momentum and angle for various types of modes, with zero torque, with a more detailed plot elsewhere [11]. Three of the distinct $\phi^*, \eta^*$ fixed points are indicated by arrows. We show a closed zero-state trajectory, labelled as (a), enclosing one of the *x*-axis fixed points $\eta^* = 0, \phi^* = \pm 2n\pi$, $n = 0, 1, 2...$, with equal portions on the positive and negative momentum sides. As the energy increases (from greater initial tilt angle/momentum, or from increased $\Lambda$, which is like the inverse-mass) the trajectories will expand horizontally and eventually develop a cusp at odd values of $\phi/\pi$. This corresponds to the well-known 'separatrix', with the pendulum just reaching the inverted position [19]. Beyond that are two running states (b), with the horizontal rippling trajectories seen to cover all phase values. Again, in terms of the physical BEC system, the onset of nonzero rotational momentum of the pendulum corresponds to the onset of a self-trapped number imbalance. These trajectories of the momentum-shortened BJJ pendulum are just modified versions of the fixed-length SJJ pendulum [19].

However, there are also novel BJJ modes. Figure 2 also shows (c): vertically oriented long ellipses enclosing the fixed-points $\phi^* = \pm(2n+1)\pi$, for $\eta^* = 0$ and equally positive and negative in momentum. These are the $\pi$-state oscillations, which can be accessed with suitable $\phi(0), \eta(0)$ initial values/energies and for $\Lambda < 1$. As $\Lambda$ increases, the phase-space trajectories develop a 'waist', that then has a pinch-off at the fixed point values $\phi^* = \pm(2n+1)\pi, \eta^* = 0$. This limiting trajectory is another novel 'separatrix'. The frequency $\omega_\pi$ dips to zero, and there is a symmetry-breaking onset of $\pi$-state rotations, with nonzero average momentum of either sign. These are shown in figure 2 as (d): small elliptical trajectories in phase space, with $|\langle\eta\rangle|$ lying *below* the nonzero $|\eta^*|$ fixed point values marked by an arrow (in fact four such loops appear in the figure). In terms of the physical BEC system, this is an MQST state with a self-maintained number imbalance $\langle\eta\rangle$, and an average phase difference of $\pi$.

As $\Lambda$ is increased further, the excursions about $\phi^* = \pi$ reduce in amplitude and there is a flattening of oscillations in $\eta$. There is an onset of (d'): a second type of $\pi$-state rotation, of average momentum magnitude *above* the fixed point value, as seen in figure 2. (There can also exist, high-$p_\phi$ running states going over all phase values, and moving through such fixed points at odd values of $\phi/\pi$ [11].) In the physical BEC system, the fixed point separating the two types of MQST $\pi$-state is a frozen critical population imbalance $\eta^*$ with phase difference of $\pi$. In the pendulum picture, it is peculiar – an upside-down pendulum with a $\pi$-state rotation loop as in figure 1d shrunk to zero, like a hovering bee tethered by a string.

Figures 3a–d show the time variation of the fractional number imbalance (solid line) and the phase difference (dashed line). The three MQST states and three $\pi$-states all have nonzero average values of $\eta$ and $\phi$, respectively. As MQST sets in, the average imbalance $\langle\eta\rangle$ rises smoothly from zero as $\Lambda$ moves through a critical value, in a second-order type transition [11]. The inverse period for all (anharmonic) BJJ oscillations can be written in terms of elliptic functions, and in terms of scaled variables. This results in a universal curve for the dips to zero at an MQST onset [11]. Note that not all $\pi$-states are self-trapped, and not all MQST states have a nonzero average phase of $\pi$. These locally stable MQST and $\pi$ states do not occur for ideal Bose gas tunnel junctions [7]: they arise from the nonlinear interaction in the GPE.

## 4. Other systems

We now comment on extensions, and other models. First we consider relaxing the $\Phi_{1,2}(\vec{r})$ rigid profile constraint. For broad double wells, the condensate profiles could conduct rocking and breathing collective oscillations in each well of low enough frequency to mix with the tunneling modes, but these effects would be weak as interwell overlaps are small.

This mixing can also occur in another system, namely a *two*-component condensate of atoms in different hyperfine states, in a *single* harmonic trap, with a near-resonant laser kept switched on to populate the excited level [13,17]. The two-state 'tunneling' now corresponds to condensate atoms transiting between the two atomic levels $1,2$.

The use of Gaussian variational wave functions for $\Phi_{1,2}(r,t)$ (with time-varying centering and width parameters to describe rocking and breathing collective modes) results in a richer set of dynamical equations. Each of the two atomic states sees a slightly shifted harmonic potential, and their large and oscillating wave function overlap, making the parameters effectively time-dependent. The pendulum is now not only momentum-shortened, but acquires new canonical coordinates that affect its length, and corresponding new canonical momenta. (With these extra degrees of freedom, the collective modes correspond to e.g. the pendulum at zero tilt angle, but oscillating up and down in length.) However, the previous fixed-point classification of modes remain useful. The tunneling modes found previously remain valid for small amplitudes, but are shifted in frequency, and induce characteristic collective oscillations in the condensate profiles, with quantum phases thus affecting the classical variables like BEC cloud centerings [13].

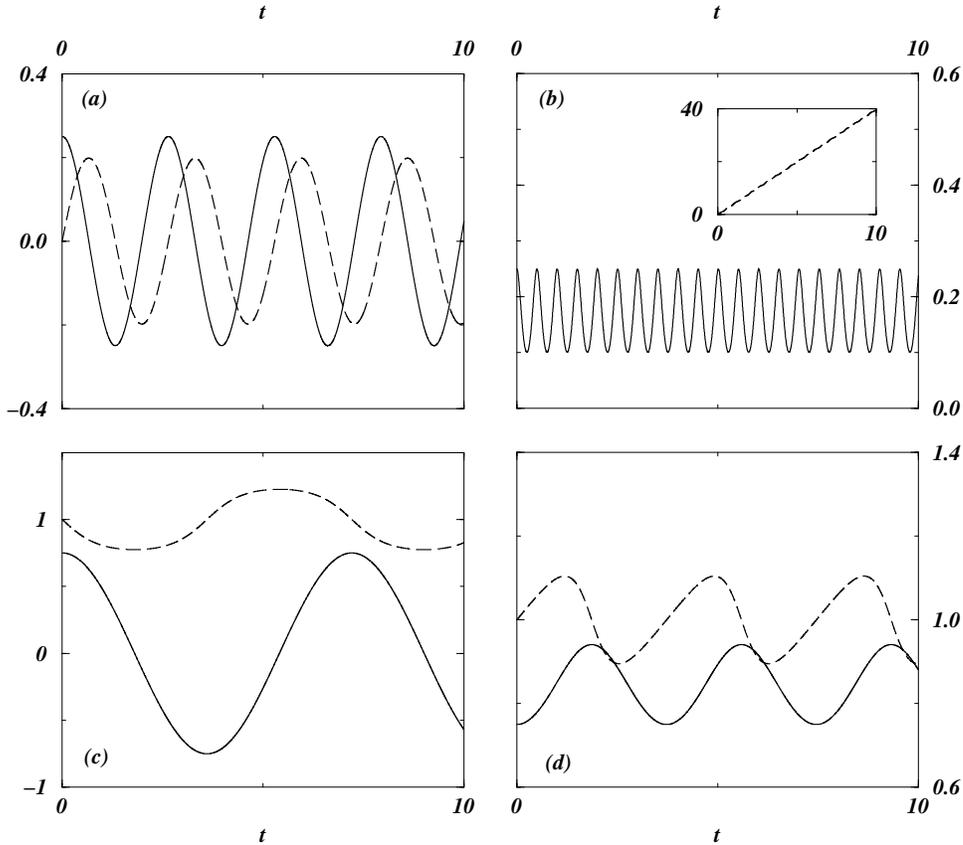

**Figure 3.** Plots versus scaled time $t/\tau$, of the BJJ number imbalance $\eta$ (solid line) and phase difference $\phi/\pi$ (dashed line) for: (**a**) zero-state oscillations with time average $\langle\eta\rangle = 0 = \langle\phi\rangle$; (**b**) running-state rotations with $\langle\eta\rangle$ nonzero and $\langle\phi\rangle \propto t$, in inset; (**c**) $\pi$-state oscillations with $\langle\eta\rangle = 0$ and $\langle\phi\rangle = \pi$; (**d**) $\pi$-state rotations of the first kind, with $\langle\eta\rangle$ nonzero and $\langle\phi\rangle = \pi$. Similar states of the second kind (see discussion in text of figure 2), are qualitatively the same, and are not shown.

The dynamics is of a BJJ-like form as in (9) and (10), rather than the form (17) for noninteracting two-level atoms. Thus the phase-coherent transit rate of two-state atoms interacting with each other in a harmonic trap is not merely a $\Lambda$-shifted version of a single ($\Lambda = 0$) Rabi frequency. The BJJ modes are multiple, and dependent on $\Lambda$: they show a frequency splitting between the zero-state and $\pi$-state, and dips of frequencies to zero at onsets to MQST states, all signatures of qualitatively new *interaction*-induced behaviour [11].

Finally, another system where these ideas and concepts are useful is the He$^3$ B superfluid [27] in two containers closed by flexible membranes that drive pairs through $4225 = 65 \times 65$ identical parallel micropores connecting the baths [28]. The pore diameter is smaller than the BCS coherence length $\xi = 650$ Å. Tunneling occurs with an ac Josephson effect at a frequency proportional to the pressure. This is the long-sought-after superfluid analogue of the superconductor Josephson junction [29], but in *s*-wave order parameter He$^3$ B, rather than in He$^4$ (where the healing length is $\xi = 1$ Å).

Oscillations with a phase difference of $\pi$ have been reported, with positive slopes in current-phase plots around $\phi = \pi$ [28]. However to explain this, one cannot just carry over the BEC ideas naively by taking $N_{1,2}$ as the entire number of Cooper pairs in the baths 1, 2: they are of the order of Avogadro's number. A fractional imbalance of just a few per cent corresponds to a tunneling current many orders of magnitude above what is observed. If the variable $\eta$ enters, it must be a fraction of some much smaller number of helium atoms; furthermore, one needs to understand why a submicron pore of cross-section $A_p < \xi^2$ can act as a tunneling barrier for a superfluid [14].

Both these issues are resolved in the Ginzburg–Landau approach by remembering that the helium wave function has to vanish at the container walls. (For a superconductor, the gradient of the wave function vanishes at surfaces.) This boundary condition means that the enforced spatial variations across the pore cross-section $A_p$ force the transverse squared-gradient terms in a GL free energy to be large and positive [14]. Averaging over transverse directions yields an effective 1D GL model, with these transverse 'kinetic energy' terms yielding an energy barrier in a region in and around the pore. This naturally generates, from overlapping wave functions, a Josephson-like phase-dependent coupling in the GL free energy.

The barrier reduces the superfluid density inside the pore, and continuity demands a depressed wave function even outside. The weakened superfluidity extends in 'overhangs' just outside the two pore openings in the two baths, that are of equal volume $A_p\xi$ in equilibrium. The wave functions in these regions decay towards the pore on one side, and must match smoothly on the other side onto flat and rigid bulk wave functions. The displacement of far-off membrane walls [28] shifts the whole wave function, and Cooper pairs are transmitted to/from the overhang regions. The entire region of depressed-superfluid density forms a weak link for phase-controlled oscillatory transport between the baths. The pair-number increments arriving and leaving on the two sides cause fractional volume increments (negative and positive) in the overhang volumes. This can be visualized as the motion of a 'Josephson piston' of a fractional displacement $\eta$ relative to its equilibrium position. The Josephson piston motion transports atoms back and forth, and the permitted piston velocity $\dot{\eta}$ is governed by the sine of the relative phase. Thus the fraction $\eta$ entering the model [14] is not a fraction of the total number of bath atoms, but of the *much* smaller number in a volume $A_p\xi$ of the 'overhangs' on either side of the pore.

The dynamics of the system is obtained by using the Josephson phase relation to link the phase rate of change to the chemical potential difference, that is a number derivative of the GL free energy. The continuity equation for matter conservation, with the GL expression for the current density, generates the second equation. Spatially integrating these dynamical equations over each bath yields equations for the fraction of atoms that tunnel, and for the phase difference across the junction. Remarkably, these $\dot{\eta}$ and $\dot{\phi}$ equations [14] are essentially those of (9) and (10). (The only complication is that the phase difference across the Josephson region is related to, but not the same as, the phase difference across the bulk bath wave functions [30].) With a single chosen parameter, the frequency magnitudes, current-phase relation, and $\omega_\pi(T)$ temperature dependences (that come from thermal phase fluctuations near the BCS transition) [14] are similiar to the experiment [28]. Thus He$^3$ B tunneling through micropores, although governed by its own characteristic physics, is also finally described by a BJJ-like dynamics.

In our model, we have assumed that the wide separation between the (submicron) pore size/coherence length and the (submillimeter) textural scale [27] justified an averaging over tensorial factors, to get a GL free energy for an effective scalar order parameter. Other approaches [31] have taken a texture-focussed view point, minimizing static free energies, and accounting for the zero-state and $\pi$-state in terms of relative orientations of the two rotation axes $\hat{n}_{1,2}$ in the two baths: parallel, or antiparallel. The two approaches emphasize different length scales, and their possibly complementary role could be explored further.

Throughout we have considered the Hamiltonian models, without damping. For double-well systems, quasiparticle damping [9] related to the chemical potential difference is probably appropriate, and a damping term $\propto -\dot{\phi}$ can be added to the r.h.s. of (9) and effects on modes studied [12]. For two-component condensates, the two-level decays will be related to the population difference, and a damping term $\propto -\eta$ could be added to (9). Such a term can surprisingly have a stabilizing effect on $\pi$-states as will be examined elsewhere.

## 5. Conclusion

We have outlined the work on the tunneling of Bose condensate atoms between double wells of an atom trap. This two-state description in terms of the relative quantum phase and the relative number difference can be mapped onto the Hamiltonian for a momentum-shortened pendulum. There is a rich set of tunneling modes, that differ from the oscillation modes for a superconductor Josephson junction. The model dynamics has several physical realizations, including single-component condensates in a double well, double-component condensates in a single well, and liquid He$^3$B tunneling through micropores.

## Acknowledgements

I am grateful to A Trombettoni for help with the figures. The work summarized and cited here resulted from a fruitful collaboration with several faculty and postdoctoral colleagues, who are gratefully acknowledged and thanked: U Al-Khawaja, M Benakli, S Fantoni, S Giovanazzi, I Marino, S Raghavan, A Smerzi and A Trombettoni. Thanks for general inspiration over the years are also due to Prof. Narendra Kumar, who, in his infectious enthusiasm to communicate the joy of physics, is absolutely unique. (In fact, there is a paradoxical Uniqueness Theorem, that says there can exist only one Kumar, and so there cannot be N Kumars...) It is a great pleasure to join colleagues in expressing through this *Pramana* special issue, our special respect and affection for the Indian physics community's most youthful stalwart.